\documentclass[12pt]{article}

\usepackage{times}
\usepackage{epsfig,amsmath}
\usepackage{subfigure}
\usepackage{graphicx}
\usepackage{color}
\usepackage{epstopdf}
\usepackage{multirow}

\newenvironment{sciabstract}{%
\begin{quote} \bf}
{\end{quote}}

\newcounter{lastnote}

\title{Experimental Test of Tracking the King Problem}

\author
{Cheng-Qiu~Hu,$^{1,2}$ Jun~Gao,$^{1,3}$ Lu-Feng~Qiao,$^{1,2}$ Ruo-Jing~Ren,$^{1,2}$ Zhu~Cao,$^{4}$ \\
Zeng-Quan~Yan,$^{1,2}$ Zhi-Qiang~Jiao,$^{1,2}$ Hao~Tang,$^{1,2}$ Zhi-Hao
 Ma,$^{5}$\\
 Xian-Min Jin$^{1,2,\ast}$\\
\
\\
\normalsize{$^1$School of Physics and Astronomy, Shanghai Jiao Tong University, Shanghai}\\ \normalsize{200240, China}\\
\normalsize{$^2$Synergetic Innovation Center of Quantum Information and Quantum Physics,}\\
\normalsize{ University of Science and Technology of China, Hefei, Anhui 230026, China}\\
\normalsize{$^3$Institute for Quantum Science and Engineering and Department of Physics,}\\ \normalsize{Southern University of Science and Technology, Shenzhen 518055, China}\\
\normalsize{$^4$Center for Quantum Information, Institute for Interdisciplinary }\\
\normalsize{Information Sciences, Tsinghua University, Beijing, China}\\
\normalsize{$^5$School of Mathematical Sciences, Shanghai Jiao Tong University,}\\
\normalsize{Shanghai 200240, China}\\
\normalsize{$^\ast$E-mail: xianmin.jin@sjtu.edu.cn}\\
}

\date{}

\begin{document}
\baselineskip24pt

\maketitle
\begin{sciabstract}
In quantum theory, the retrodiction problem is not as clear as its classical counterpart because of the uncertainty principle of quantum mechanics. In classical physics, the measurement outcomes of the present state can be used directly for predicting the future events and inferring the past events which is known as retrodiction. However, as a probabilistic theory, quantum-mechanical retrodiction is a nontrivial problem that has been investigated for a long time, of which the Mean King Problem is one of the most extensively studied issues. Here, we present the first experimental test of a variant of the Mean King Problem, which has a more stringent regulation and is termed ``Tracking the King". We demonstrate that Alice, by harnessing the shared entanglement and controlled-not gate, can successfully retrodict the choice of King's measurement without knowing any measurement outcome. Our results also provide a counterintuitive quantum communication to deliver information hidden in the choice of measurement.\\
\end{sciabstract}

The thought experiment, named the Mean King Problem (MKP), is a quantum-mechanical retrodiction problem, which originates from the paper in 1987 by Vaidman, Aharonov and Albert\cite{vaidman1987ascertain}. The original research dealt with spin-half particles, i.e., the case in 2-dimension Hilbert space, and was generalized to prime dimensionality\cite{englert2001mean} and power of prime dimensionality\cite{aravind2003solution,durt2010mutually}. Actually, the MKP and its variants are continuously developed in the general frame of quantum theory\cite{revzen2013maximal,kalev2014quantum,assad2014encoding,englert2001universal}. 

{In the original version described by Vaidman, Aharonov and Albert, a measurement  chosen from \{$\sigma_{x}, \sigma_{y}, \sigma_{z}$\} is performed on a spin-$\frac{1}{2}$ particle at a given time. By means of measurements carried out both before and after the time in question, they came up with a method to ascertain  the result of the spin measurement, even though they do not know in which direction the spin was measured and no matter what those measured results may happen to be. The solution of this problem is intriguing because it is by no means obvious to make definite inferences about $\sigma_{x}, \sigma_{y}, \sigma_{z}$.}

In the Mean King Problem, a physicist named Alice prepares a two prime-dimensional particles state and avails one of the particles to the King who then measures his particle by freely choosing one of the mutually unbiased bases (MUB)\cite{wootters1989optimal,bandyopadhyay2002new,vourdas2004quantum,gibbons2004discrete,durt2010mutually}. The King challenges Alice to perform a control measurement of her liking and then state the outcome of his measurement correctly with the knowledge of his choice of measurement. 

It seems that in the MKP the physicist can assign the state of the system in the past. However, with the knowledge of the King's choice of measurement, this state retrodiction is actually conditional. What if the King doesn't tell Alice his choice? This question leads to another extension of the MKP termed ``Tracking the King Problem"(TKP). In this new problem, Alice is not informed the King's choice (nor the outcome). Inversely her control measurement is designed to track the basis that the King used. The interesting thing is, in TKP, the King's choice of measurement (without recording the outcome) can be viewed as a novel quantum communication \cite{bennett1984update,bennett1992communication} signal, viz. a message sent to Alice\cite{kalev2013choice}.

The counterintuitive part of this problem is that in classical physics, measurements without outcomes being recorded carry no information and hence cannot be used for communication. However, such a measurement in quantum mechanics causes traceable disturbance to the measured system so that Alice's control measurement can retrieve the King's choice. Until recently, this thought experiment has not been realized yet. 

In this work, we experimentally demonstrate the TKP of the 2-dimension case in an optical system. By sharing maximally-entangled photon pairs and employing a controlled-not gate (C-NOT) as Alice's control measurement, we show that it is possible for Alice to retrodict the choice of King's measurement without knowing any measurement outcome. The realization of this thought experiment unravels the intrinsic characteristics of quantum measurement and deepens our understanding of quantum theory. Furthermore, the choices of measurements can also be seen as the delivered messages, which put it forward to realize a counterintuitive quantum communication.

The scheme of the TKP is shown in Fig. 1. Alice starts with preparing one of the maximally-entangled bipartite states:%
\begin{equation}
\left\vert c,r;s\right\rangle _{1,2}=\frac{1}{\sqrt{d}}\sum%
\limits_{n=0}^{d-1}\left\vert n\right\rangle _{1}\left\vert c-n\right\rangle
_{2}\omega ^{sn^{2}-2rn}
\end{equation}
where $d=2$; $\omega =i$; $c,r,s=0,1,\ldots ,d-1$; and $\left\vert
n\right\rangle =\left\vert n+d\right\rangle $ for any $n$. The subscript $s$
labels a basis for Hilbert space of the two qubits, while $c$, $r$ label the 
$d^{2}$ orthonormal states within the basis. For example, $s=c=r=0$
represents the state of $\left\vert \phi ^{+}\right\rangle =\frac{1}{\sqrt{2}%
}\left( \left\vert 1\right\rangle \left\vert 1\right\rangle +\left\vert
0\right\rangle \left\vert 0\right\rangle \right) $. 
{\section*{Experimental Setup}}
The schematic view of the experimental setup is shown in Fig. 2. In Alice's station,
polarization-entangled photon pairs are prepared via type-II spontaneous
parametric down-conversion. The 390nm pulsed laser is generated in the LiB$%
_{3}$O$_{5}$ (LBO) crystal by frequency doubling, and is focused on the 2mm-thick
type-II degenerate non-colinear cut beta-barium-borate (BBO) crystal as a
pump light\cite{kwiat1995new}. The combination of a half-wave plate (HWP) and a 1mm-thick BBO crystal is used in each arm to compensate the spatial and temporal walk-off of the generated photons. Two extra HWPs are used here to create four Bell states as different initial states:%
\begin{eqnarray}
\left\vert \phi ^{\pm}\right\rangle  &=&\left\vert H\right\rangle
_{1}\left\vert H\right\rangle _{2}\pm\left\vert V\right\rangle _{1}\left\vert
V\right\rangle _{2}  \nonumber \\
\left\vert \psi ^{\pm}\right\rangle  &=&\left\vert H\right\rangle
_{1}\left\vert V\right\rangle _{2}\pm\left\vert V\right\rangle _{1}\left\vert
H\right\rangle _{2}  \
\end{eqnarray}
where $H$ ($V$) represents horizontal (vertical) polarization of the qubit.

One of the entangled photons (P1) is stored with a fixed time delay by being coupled into a single-mode fiber. The other one (P2) is sent to the King's station, where a standard single-qubit analyzer consisting of two quarter-wave plates (QWPs) and a polarizer (POL) serves as the required nonselective measurement (see the black box in Fig. 2.). The collapsed photon state is then coupled into a single-mode fiber and sent back into the C-NOT gate together with the P1 in Alice station. We measure a truth table in the computational basis $ZZ$ and obtain an average fidelity of the C-NOT gate up to 0.827. More details about the implementation and optimization of C-NOT gate can be found in Method and Fig. 3. With single-qubit polarization projection for each photon and coincidence measurement, we are able to project P1 and P2 onto four Bell states. To test such an ability for distinguishing Bell states, we inject all the four maximally entangled states in Eq. 3 and measure their coincidence, in the basis \{$D,A$\} for the control arm and in the basis \{$H,V$\} for the target arm. The results shown in Fig. 3(c) shows a good ability of transforming the maximally entangled states to corresponding product states, which is directly related to the performance of the required control measurement.
{\section*{Results}}
We test the TKP in two different initial conditions, in which Alice prepares the singlet state $\left\vert\psi^{-}\right\rangle$ and the triplet state $\left\vert\phi^{+}\right\rangle$ respectively. After receiving the qubit sent by Alice, the King performs the nonselective measurement chosen from \{$\sigma_{x}, \sigma_{y}, \sigma_{z}$\} by changing the angles of the combination of the POL and the QWPs. Then he sends the qubit back to Alice. The reunited two qubits go through the device consisting of two QWPs and a HWP in order to compensate the polarization rotation induced by the fibers. Afterwards, Alice performs the control measurement in the Bell basis \{$\left\vert\phi^{+}\right\rangle, \left\vert\phi^{-}\right\rangle, \left\vert\psi^{+}\right\rangle, \left\vert\psi^{-}\right\rangle$\} on the two qubits to retrieve the King's choice of measurement $b$. The expected probabilities are listed in Table \uppercase\expandafter{\romannumeral1} as a theoretical truth table.
\begin{table*} \centering%
\renewcommand{\tablename}{Table.}
\caption{\textbf{The theoretical truth table for initial states $\left\vert \phi ^{+}\right\rangle $ and $\left\vert \psi ^{-}\right\rangle $}}\label{TableKey}%
\begin{tabular}{c|cccc|cccc}
\hline\hline
\multirow{2}{*}{\ The King's Choice\ } & \multicolumn{8}{|c}{Alice's outcomes} \\ \cline{2-9}
& \multicolumn{4}{|c|}{initial state $\left\vert \phi ^{+}\right\rangle $} & 
\multicolumn{4}{|c}{initial state $\left\vert \psi ^{-}\right\rangle $}
\\ \hline
$b$ & $DH$ & $DV$ & $AH$ & $AV$ & $DH$ & $DV$ & $AH$ & $AV$ \\ \hline
$\sigma _{x}$ & 0.5 & 0.5 & 0 & 0 & 0 & 0 & 0.5 & 0.5 \\ 
$\sigma _{y}$ & 0 & 0.5 & 0.5 & 0 & 0 & 0.5 & 0.5 & 0 \\ 
$\sigma _{z}$ & 0 & 0.5 & 0 & 0.5 & 0.5 & 0 & 0.5 & 0 \\ \hline\hline
\end{tabular}%
\end{table*}%


The experimental results obtained from Alice's control measurement are shown in Fig. 4, from which we do observe a good agreement with the theoretical truth table. Both the expected and unexpected probabilities are much far away from 0.25, an uniform probability distribution. It should be noticed that the King's measurement outcomes are not used at all in the control measurement in our experiment, but are only borrowed to check whether our results are consistent with the theoretical truth table. The reliability of identifying the King's choice, the number of successful events divided by all trial times, is found to be up to 0.813 on average, well going beyond the rate of 0.5 from wild guessing.  

The demonstrated abilities of retrodicting the King's choice of measurement apparently can be considered as a form of communication protocol. We ask two volunteer students to act as Alice and the King, who execute the whole process like a game. In the case that Alice prepares the initial state $\left\vert\phi^{+}\right\rangle$, the King randomly choose a series of nonselective measurements $b$ out of \{$\overset{..}{0},0,1$\}. Alice then can ``guess" the King's choice from \{$\overset{..}{0},0,1$\} relying on her measurements and the truth table. We show a fraction of trial events in Fig. 5. We can see that, while the reliability is not unit, an appropriate statistics can help Alice unambiguously reveal the King's choice and win the game. 

In summary, by experimentally testing a variant of MKP, i.e., the TKP, we exemplify how tasks that seem impossible by classical reasoning can be experimentally achieved within quantum mechanics frameworks. By harnessing the shared entanglement and controlled-not gate, we demonstrate that we can successfully retrodict the choice of King's measurement without knowing any measurement outcome. Our results provide a strong distinction for the features between classical and quantum systems. While performing nonselective measurements on classical systems, no matter how correlated they are, cannot carry or manipulate information \cite{schwinger2001quantum,diosi2011short}, the trackability of nonselective measurements on quantum systems inevitably introduces distinguishable disturbance \cite{kalev2013choice}. 

The realization of such a thought experiment also provides a counterintuitive quantum communication to deliver information hidden in the choice of measurement: the King sends a message that he doesn't want to send. We may also conceive an anti-eavesdropping scheme to reveal all eavesdroppers' actions when they try to hack into an entanglement-distributed quantum network: tracking the King.

\section*{Methods}
\textbf{Experimental Details:} The C-NOT gate, as a key part of the control measurement, is realized by an essential partial polarization beam splitter (PPBS-\uppercase\expandafter{\romannumeral1}) and two supplemental PPBS-\uppercase\expandafter{\romannumeral2}s\cite{ralph2002linear,okamoto2005demonstration}. The PPBS-\uppercase\expandafter{\romannumeral1} reflects vertically polarized light perfectly and reflects (transmits) 1/3 (2/3) of horizontally polarized light, which performs as a quantum phase gate while attenuates the $H$ components by a factor of $1/\sqrt{3}$. The two PPBS-\uppercase\expandafter{\romannumeral2}s are inserted to each of the interferometer paths as local polarization compensators, which transmits $H$ components perfectly and transmits (reflects) 1/3 (2/3) $V$ components. The two incident photons are ensured to perfectly interfere on the PPBS-\uppercase\expandafter{\romannumeral1} to erase their which-way information. We optimize their spatial and temporal overlap by observing the Hong-Ou-Mandel interference with an injected identical polarization of $H$. Fig. 3(a) shows the measured Hong-Ou-Mandel dip, with which we obtain a visibility of 66.3\% by fitting with the Gaussian curve. In light of the ideal value of 80\% determined by the specification of the PPBS-\uppercase\expandafter{\romannumeral1}, the achieved visibility represents a good mode match in both space and time. As is shown in Fig. 5(b), we measure a truth table in the computational basis $ZZ$ and obtain an average fidelity of the C-NOT gate up to 0.827. {Take it specifically, if we inject the $V$ into the control arm, the target bit will keep its original state. Otherwise, if we inject $H$ into the control arm, the target bit will flip,  which we denote using red bars in Fig. 3(b). With this basic function, we can use the C-NOT gate to discriminate four Bell states. For example, the injection of state $\left\vert\phi^{+}\right\rangle$ will come out with the result $D$ in the control arm and $V$ in the target arm (see Table. 2).}
\begin{table} \centering%
\renewcommand{\tablename}{{Table.}}
\caption{{\textbf{Discrimination of different Bell states with C-NOT gate}}}\label{TableKey}%
\begin{tabular}{c|c|c|c|c}
\hline \hline 
Injection states &$\left\vert\phi^{+}\right\rangle$ & $\left\vert\phi^{-}\right\rangle$ & $\left\vert\psi^{+}\right\rangle$ & $\left\vert\psi^{-}\right\rangle$\\
\hline  
Control bit& $D$ & $A$ & $D$ & $A$\\
\hline 
Target bit& $V$ & $V$ & $H$ & $H$\\
\hline 
\hline
\end{tabular}
\end{table}
%

\noindent \textbf{Decoding the King's Message:} Alice keeps qubit 1
(labeled by the subscript 1) and sends qubit 2 to the King. Then the King chooses one of the $d+1$ MUBs as his nonselective measurement
on the qubit he received. His choice is labeled by $b=\overset{..}{0}%
,0,1,\ldots ,d-1$:%
\[
\left\{ \left\vert n\right\rangle \right\} _{n=0}^{d-1}
\]%
for $b=\overset{..}{0}$ and%
\[
\left\{ \left\vert m;b\right\rangle =\frac{1}{\sqrt{d}}\sum_{n=0}^{d-1}\left%
\vert n\right\rangle i^{bn^{2}-2mn}\right\} _{m=0}^{d-1}
\]%

for $b=0,1\ldots ,d-1$. Note that in this scheme the King's measurement
outcome is completely irrelevant, and he sends back the collapsed qubit to Alice without
recording any results afterwards. Now the two-qubit state can be expressed as follow:%
\begin{equation}
\rho _{1,2}=\sum_{m=0}^{d-1}\left\vert m;b\right\rangle _{1}\left\langle
m;b|c,r;s\right\rangle _{1,2}\left\langle c,r;s|m;b\right\rangle
_{1}\left\langle m;b\right\vert 
\end{equation}

At last, Alice measures the two-qubit system in the basis denoted by:%
\[
\left\{ \left\vert c^{\prime },r^{\prime };s\right\rangle _{1,2}\right\}
_{c^{\prime },r^{\prime }=0}^{d-1}
\]%
from which Alice can retrieve the choice of the King's measurement, i.e., the label $b$ according to the following decoding table:
\begin{eqnarray*}
c &\neq &c^{\prime }\rightarrow b=s+\frac{r-r^{\prime }}{c^{\prime }-c} \\
r &\neq &r^{\prime },c=c^{\prime }\rightarrow b=\overset{..}{0}, \\
r &=&r^{\prime },c=c^{\prime }\rightarrow inconclusive.
\end{eqnarray*}

\subsection*{Acknowledgments.}
The authors thank Jian-Wei Pan for helpful discussions and suggestions. This work was supported by National Key R\&D Program of China (2017YFA0303700); National Natural Science Foundation of China (NSFC) (61734005, 11761141014, 11690033); Science and Technology Commission of Shanghai Municipality (STCSM) (15QA1402200, 16JC1400405, 17JC1400403); Shanghai Municipal Education Commission (SMEC)(16SG09, 2017-01-07-00-02-E00049); X.-M.J. acknowledges support from the National Young 1000 Talents Plan.\\
\subsection*{Data availability}
The data that support the findings of this study are available from the corresponding author on request.

\subsection*{Competing Interests} The authors declare no competing interests.
 
\subsection*{Author contributions} X.-M.J. conceived the project and designed the experiment. C.-Q.H., J.G., L.-F.Q., R.-J.R., Z.-Q.Y., Z.-Q.J., H.T.and X.-M.J. performed the experiment. Z.C. and Z.-H.M. conducted the theoretical work. X.-M.J. and C.-Q.H. analysed the data and wrote the paper with the input from all the authors. 

\subsection*{Correspondence} Correspondence and requests for materials should be addressed to X.-M.J\\
(xianmin.jin@sjtu.edu.cn).

\clearpage
\begin{figure}
\centering
\includegraphics[width=0.7\columnwidth]{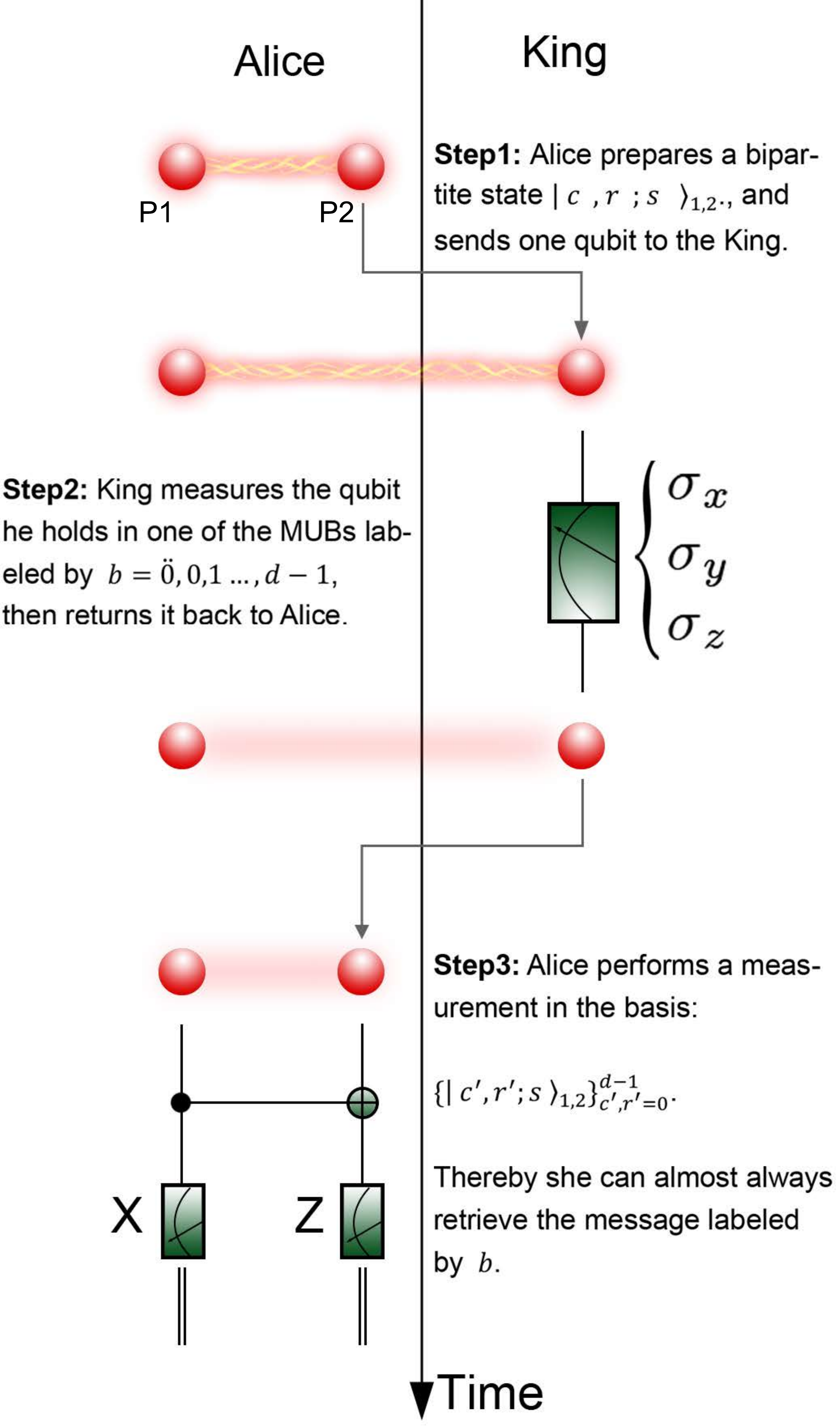}
\renewcommand{\figurename}{Fig.}
\caption{\textbf{The scheme of  ``Tracking the King Problem." }}
\label{fig1}
\end{figure}
\clearpage
\begin{figure}
\centering
\includegraphics[width=1\columnwidth]{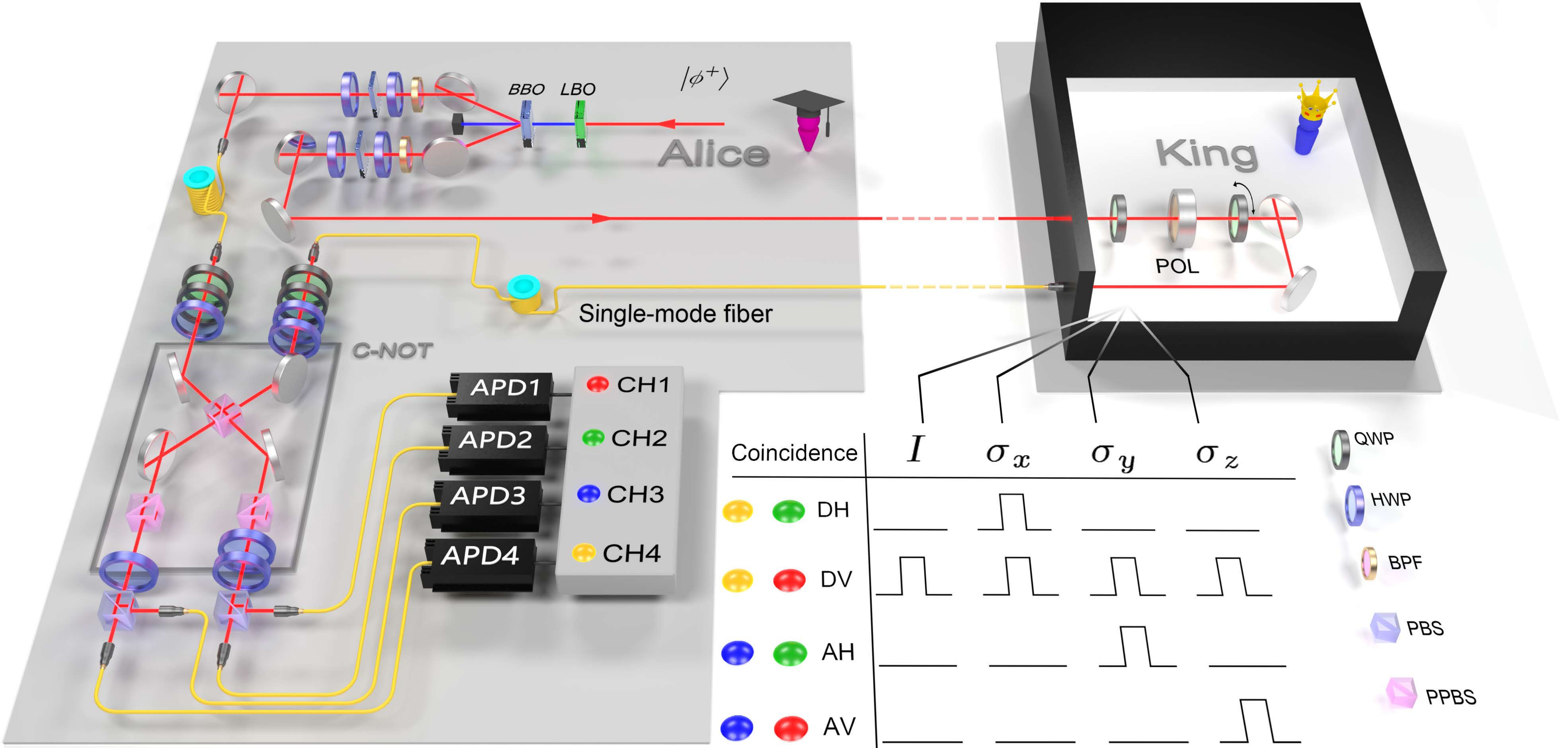}
\renewcommand{\figurename}{Fig.}
\caption{\textbf{The experimental setup. }Maximally-entangled photon states are generated via spontaneous parametric down conversion on Alice's station (see the upper left corner of the sketch). For a certain initial state (featured an illustration of $\left\vert \phi ^{+}\right\rangle $ here), The King chooses one of the MUBs to measure the qubit sent to him (in the black box) and returns it back to Alice through the single-mode fiber. In Alice's station, she can retrodict the King's choice of measurement with her control measurement (see the lower left corner of the sketch). The King's choice of nonselective measurements in MUBs ($\sigma _{x}$,$\sigma _{y}$,$\sigma _{z}$) will lead corresponding coincidences (processed by a FPGA), which is listed in the table at the bottom right corner of the figure (each coincidence click has the same probability of 0.5).The identical matrix $I$ means doing nothing to the qubit, of which
only coincidence $DV$ will click. BPF: bandpass filter. POL: polarizer. APD: avalanche single-photon detector.}
\label{fig2}
\end{figure}
\clearpage
\begin{figure}
\centering
\par

\includegraphics[width=1\columnwidth]{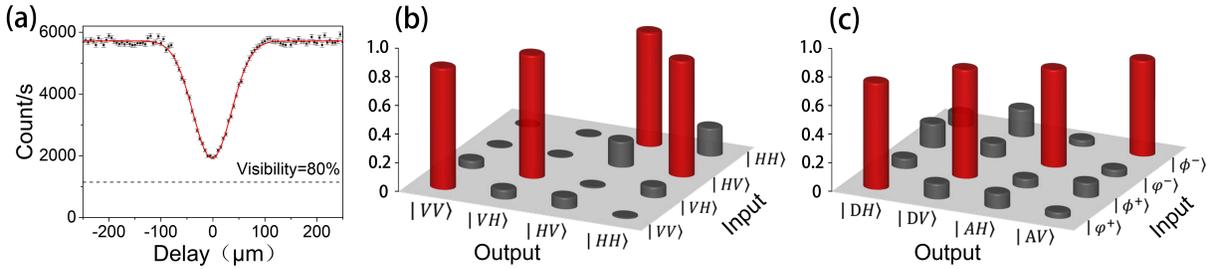}
\par
\renewcommand{\figurename}{Fig.}
\caption{\textbf{The complete characterization of the control measurement system. }\textbf{a.} By injecting the state $\left\vert H\right\rangle_{a}\left\vert H\right\rangle_{b}$, the two-photon interference is shown via the Hong-Ou-Mandel dip with the obained visibility of 66.3\%. \textbf{b.} The truth table measured in the computational basis $ZZ$. \textbf{c.} Demonstration of the ability of the C-NOT gate to transform the maximally entangled states into corresponding product states. }
\label{fig3}
\end{figure}
\clearpage
\begin{figure}
\centering
\includegraphics[width=1\columnwidth]{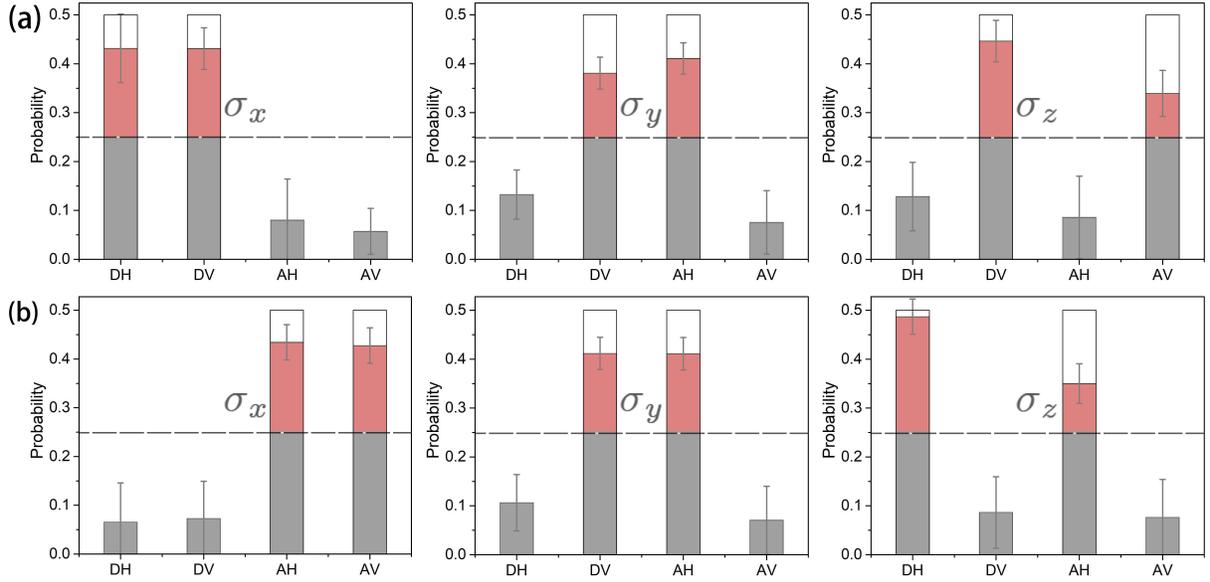}
\renewcommand{\figurename}{Fig.}
\caption{\textbf{The measured truth tables of retrieving the King's choice of measurement. }The empty histograms represent the theoretical probabilities while the color filled histograms represent the experimentally measured probabilities, from which we can obtain the average reliability up to 0.813. According to the histograms beyond the threshold line (at probability 0.25 in the case of uniform distribution), Alice can retrodict the King's choice of nonselective measurement in MUBs (indicated by $\sigma _{x}$,$\sigma _{y}$,$\sigma _{z}$ on the center of each subgraphs) referring to the Table \uppercase\expandafter{\romannumeral1}. \textbf{a.} When the initial state is $\left\vert \phi ^{+}\right\rangle$. \textbf{b.} When the initial state is $\left\vert \psi ^{-}\right\rangle$. The error bars are calculated with Poissonian statistics of the detection process taken into account. }
\label{fig4}
\end{figure}
\clearpage
\begin{figure}
\centering
\includegraphics[width=1\columnwidth]{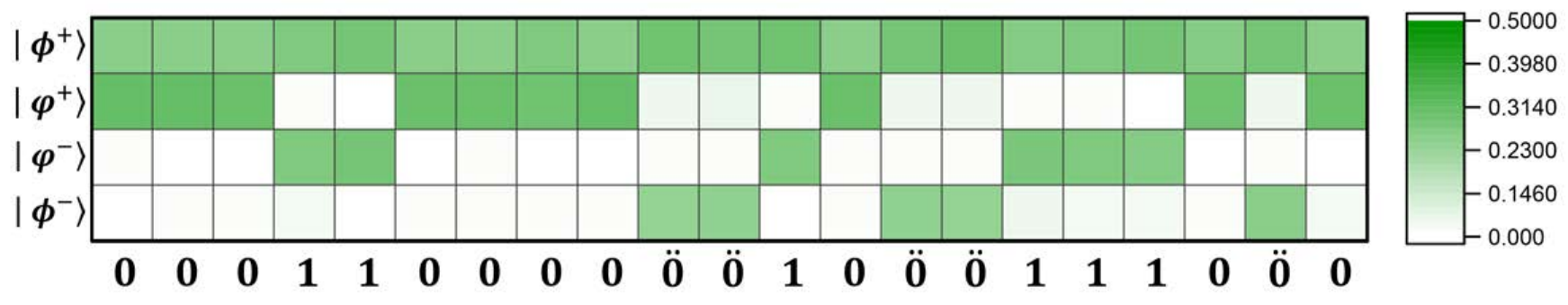}
\renewcommand{\figurename}{Fig.}
\caption{\textbf{A stream of random number obtained by retrieving the King's measurements.} In a game between Alice and the King, after the initial state $\left\vert \phi ^{+}\right\rangle$ is distributed, a series of $b$ values are randomly chosen by the King and the corresponding nonselective measurements in MUBs are implemented successively. Then Alice can ``guess" the King's choice by her own measurement. The measured probabilities are represented by the color and the identified random numbers are listed below.}
\label{fig5}
\end{figure}
\end{document}